\documentclass{article}
\usepackage{spconf,amsmath,graphicx}
\usepackage{subfigure}
\usepackage{multirow}
\usepackage{array}
\usepackage{booktabs}

\def\L{{\cal L}}

\title{AlignTTS: Efficient Feed-Forward Text-to-Speech System \\ without Explicit Alignment} 
\name{Zhen Zeng, Jianzong Wang\sthanks{Corresponding author: Jianzong Wang, jzwang@188.com}, Ning Cheng, Tian Xia, Jing Xiao }
\address{Ping An Technology (Shenzhen) Co., Ltd.}
\begin{document}
%
\maketitle
\begin{abstract}
Targeting at both high efficiency and performance, 
we propose AlignTTS to predict the mel-spectrum in parallel. 
AlignTTS is based on a Feed-Forward Transformer 
which generates mel-spectrum from a sequence of characters, 
and the duration of each character is determined by a duration predictor.
Instead of adopting the attention mechanism in Transformer TTS 
to align text to mel-spectrum, 
the alignment loss is presented to consider all possible alignments in training 
by use of dynamic programming. 
Experiments on the LJSpeech dataset show that 
our model achieves not only state-of-the-art performance 
which outperforms Transformer TTS by 0.03 in mean option score (MOS), 
but also a high efficiency which is more than 50 times faster than real-time.

\end{abstract}
\section{Introduction}
\label{sec:intro}

Recent advances in text-to-speech (TTS) are driven by the success of autoregressive neural network, 
such as Char2wav \cite{Char2wav}, VoiceLoop \cite{Voiceloop}, Tacotron \cite{Tacotron}, 
Tacotron 2 \cite{Tacotron2}, Deep Voice 1 \cite{DeepVoice1},Deep Voice 2 \cite{DeepVoice2}, 
Deep Voice 3 \cite{DeepVoice3}, Clarinet \cite{Clarinet} 
and Transformer TTS \cite{TransformerTTS}. These models employ a sequence-to-sequence (seq2seq) framework 
with attention mechanism, where the encoder converts the input text into a context-sensitive features 
and the decoder generates the mel-spectrum based on the context-sensitive features. In addition, 
a neural vocoder, such as WaveNet \cite{wavenet}, WaveRNN \cite{WaveRNN} and WaveGlow \cite{WaveGlow}, 
is usually used to synthesize waveform from the mel-spectrum. Although the 
autoregressive-based TTS system has achieved satisfactory synthesis results, low inference efficiency 
of autoregressive neural network still limits its application in a real-time dialogue system. 

Most recently, many non-autoregressive architectures for TTS are proposed to increase the speed of 
synthesis speech, such as ParaNet \cite{ParaNet} and FastSpeech \cite{FastSpeech}. 
Combining with parallel neural vocoder, these models can achieve real-time speech synthesis 
by predicting mel-spectrum in parallel. 
However, since it is difficult for these models to learn the alignment between text and mel-spectrum, 
a pre-trained auto-regressive model is required to guide their training, 
which limits them to achieve better performance. 
Therefore, in order to improve the quality of synthesis speech, it is necessary for non-autoregressive 
TTS systems to design more appropriate method to directly learn the alignment. 

Reviewing the traditional statistical parametric speech synthesis 
system based on the Hidden Markov Models (HMMs) \cite{SpeechSynthesisHMMs,taylor2009TTSBook}, 
the Baum-Welch algorithm \cite{BaumWelch} is used to estimate the parameters of HMMs if the alignment between states
and observations is unknown. In detail, the Baum-Welch algorithm considers all possible alignments and 
sums the estimated value from each alignment as the final estimated value, where each observation is assigned to 
every state in proportion to probability of that generating the observation. 
Inspired by this idea, we present a novel method to 
improve the alignment for the non-autoregressive TTS system. 

In this paper, 
we propose AlignTTS to generate the mel-spectrum in parallel, 
whose training does not require guidance from other autoregressive TTS systems. 
AlignTTS consists of a Feed-Forward Transformer, 
a duration predictor and a mix density network. 
The Feed-Forward Transformer is a feed-forward network to transform text to mel-spectrum, 
where the duration of each character predicted by the duration predictor is required to 
regulate the alignment in inference. 
Experiments on the LJSpeech dataset show that AlignTTS achieves state-of-art performance 
which outperforms Transformer TTS by a gap of 0.03 in MOS.
Meanwhile, it takes only 0.18 seconds to synthesize approximately 10 seconds of speech in our model, 
which is more than 50 times faster than real-time.
And the main contributions of our works as follows:
\begin{itemize}
\item Due to the feed-forward network structure, 
AlignTTS can generate the mel-spectrum in parallel. 
Combining with WaveGlow Vocoder, the speech synthesis speed 
is more than 50 times faster than real-time;  

\item The alignment loss is proposed to guide AlignTTS
to learn the alignment between the text and mel-spectrum. 
Specifically, the learned alignment is more precise in aligning text and mel-spectrum 
than the attention alignment from Transformer TTS \cite{TransformerTTS}, 
so that the more accurate conversion from text to mel-spectrum is learned in AlignTTS.
\end{itemize}

\section{Architecture of AlignTTS}
\label{sec:architecture}

In order to predict the mel-spectrum in parallel, we propose AlignTTS, a feed-forward network, 
which contains three modules including the Feed-Forward Transformer module, the duration 
predictor and the mix density network. Each module is described in detail in this section.

\begin{figure}[t]
  \subfigure[ Feed-Forward Transformer ]{
      \begin{minipage}[b]{0.55\linewidth} 
          \centering 
          \includegraphics[width=0.99\linewidth]{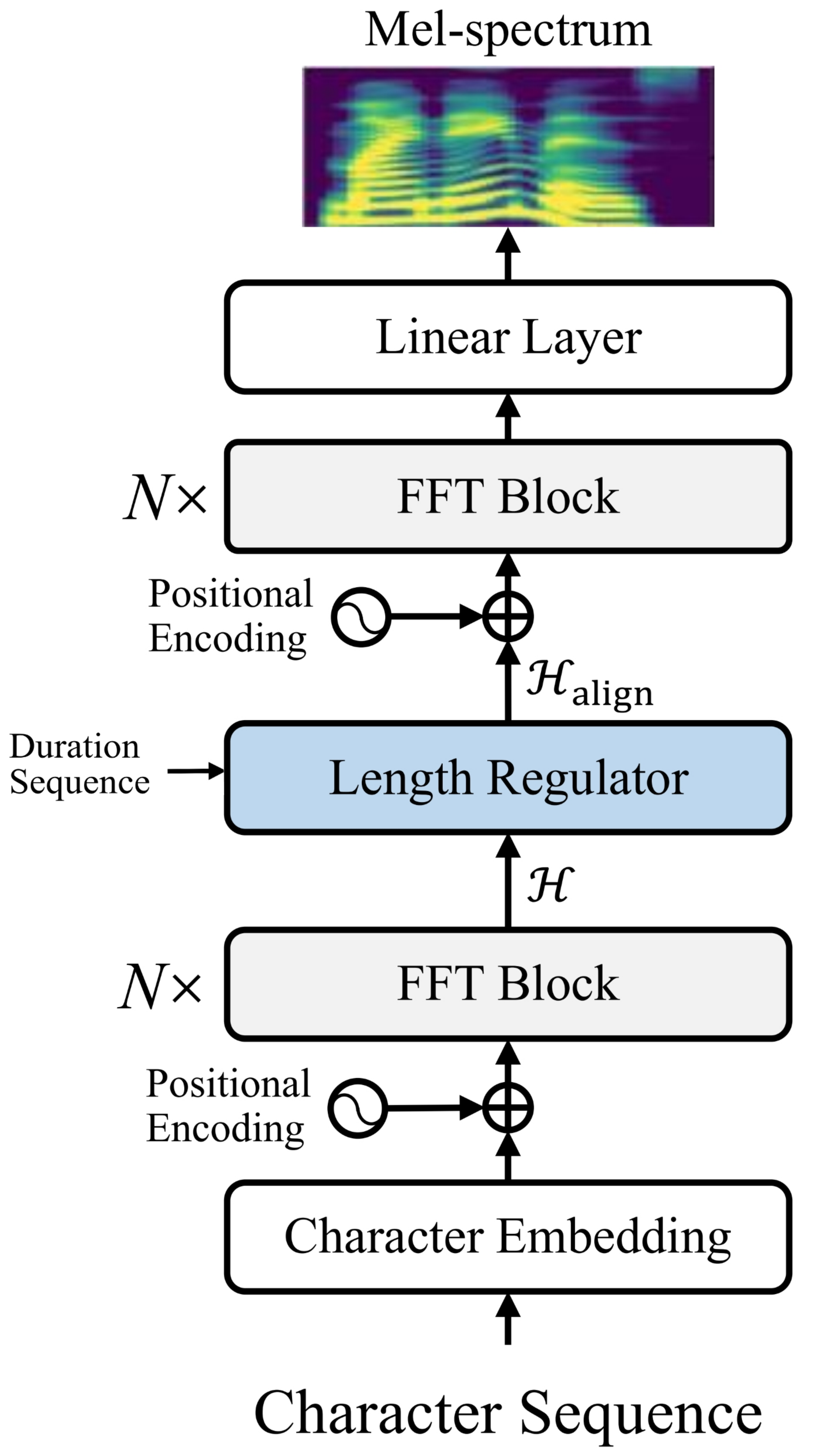}
      \end{minipage}
      \label{Figure1a}
  }
  \subfigure[ FFT Block ]{
      \begin{minipage}[b]{0.38\linewidth} 
          \centering 
          \includegraphics[width=\linewidth]{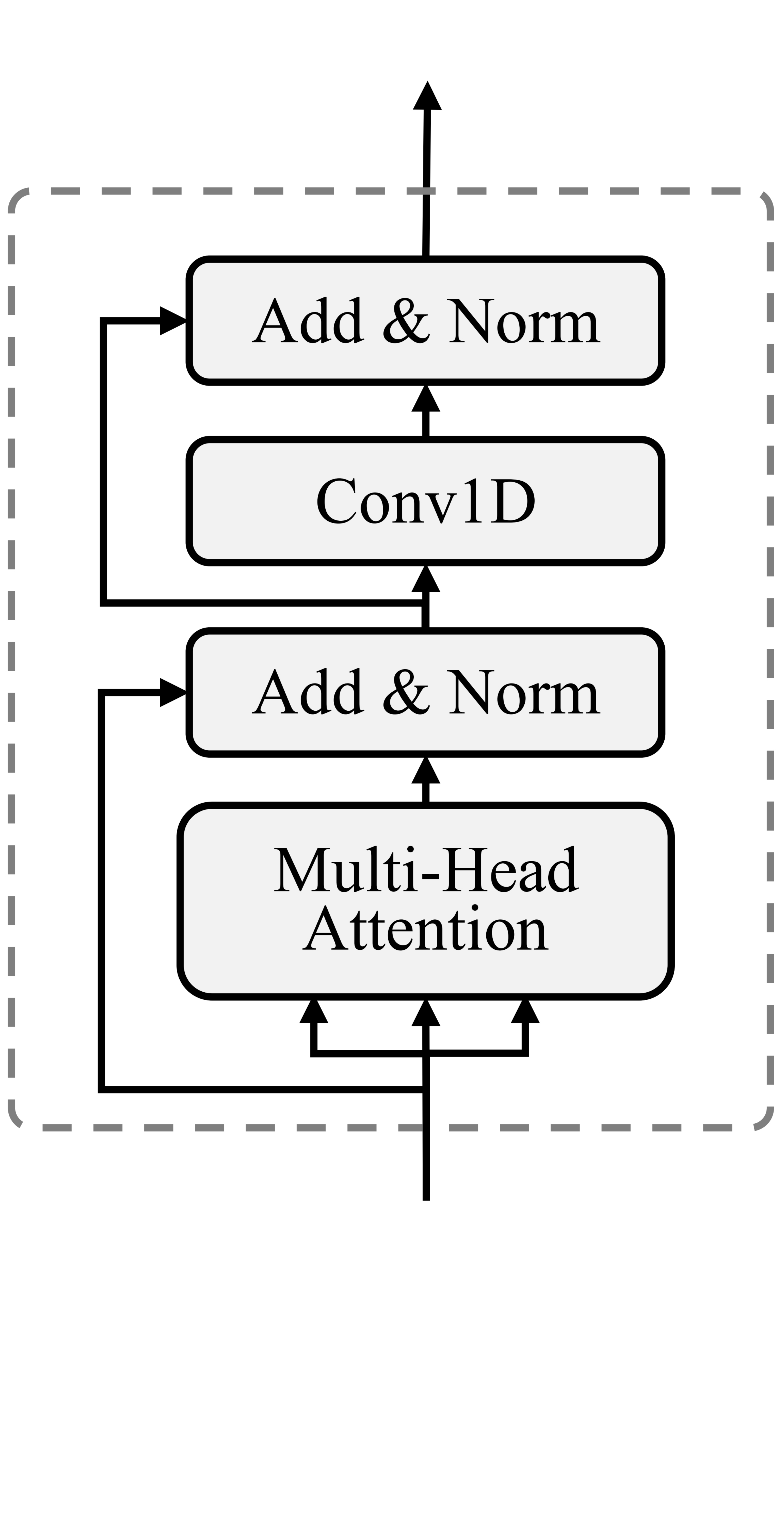}
      \end{minipage}
      \label{Figure1b}
  }
  \caption{(a) Feed-Forward Transformer for text to mel-spectrum transformation. 
  (b) FFT Block, composed of self-attention and 2-layer 1D conventional network.}
\end{figure}

\subsection{Feed-Forward Transformer}
\label{ssec:Feed-Forward Transformer}

The Feed-Forward Transformer (FFT) module 
stacks a character embedding, multiple FFT blocks, a length regulator and a linear layer for text to mel-spectrum 
transformation, as shown in Figure 1(a). Multiple FFT blocks are divided into two parts by 
the length regulator. Each FFT block is composed of a self-attention from Transformer \cite{AttentionAllYouNeed} 
and 2-layer 1D convolutional network, 
where residual connections, layer normalization and dropout are used, as shown in Figure 1(b). 
In addition, the length regulator is used to regulate the alignment between text and mel-spectrum 
according to the given duration sequence, which is generated by the duration predictor in inference. 
Since the adjustment method of the length regulator 
is the same as description in FastSpeech \cite{FastSpeech}, the voice speed and the breaks between words can 
also be controlled in our model.

\begin{figure}[t]
  \subfigure[ Duration Predictor ]{
      \begin{minipage}[b]{0.38\linewidth} 
          \centering 
          \includegraphics[width=0.99\linewidth]{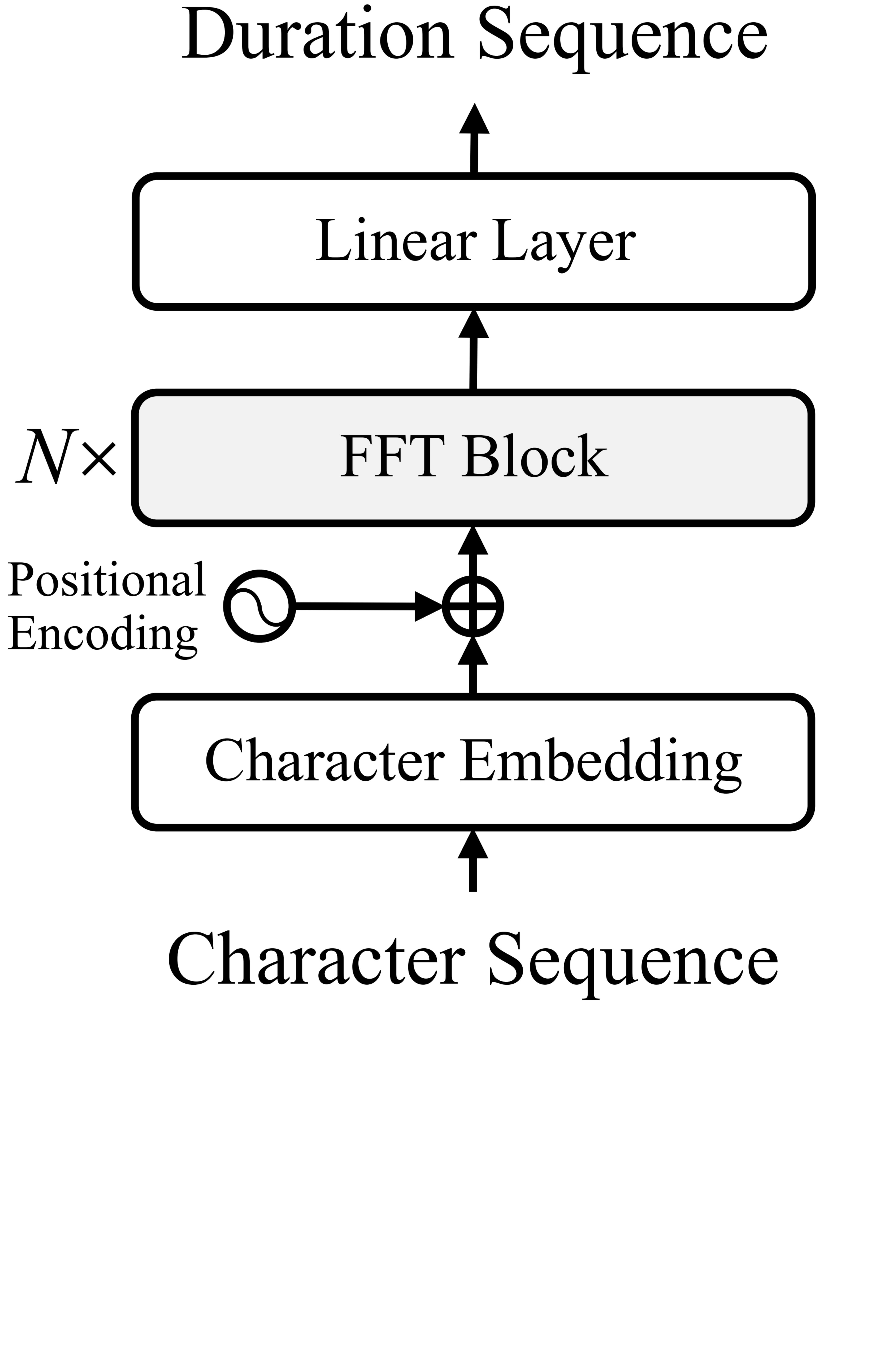}
      \end{minipage}
      \label{Figure2a}
  }
  \subfigure[ Mix Density Network ]{
      \begin{minipage}[b]{0.62\linewidth} 
          \centering 
          \includegraphics[width=\linewidth]{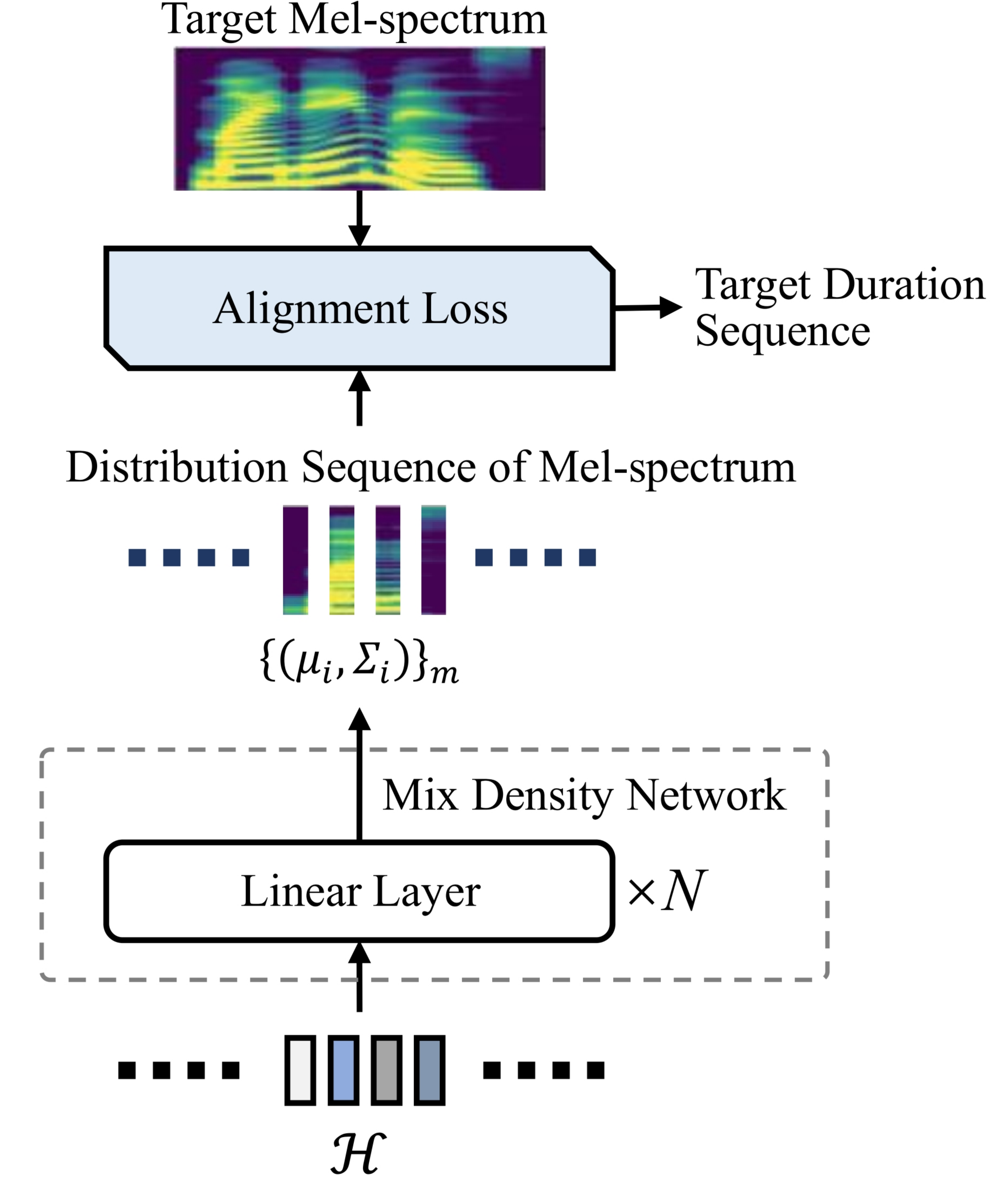}
      \end{minipage}
      \label{Figure2b}
  }
  \caption{(a) The duration predictor for predicting the duration sequence in inference.
    (b) The Mix Density Network for learning the alignment between text and mel-spectrum. }
\end{figure}

\subsection{Duration Predictor} 
\label{ssec:Duration Predictor}


Different from conventinal speech synthesis systems employing the attention mechanism, 
AlignTTS uses the duration predictor to predict the alignment between text and mel-spectrum, 
which can be calculated in parallel. 
The duration predictor consists of a character embedding layer, multiple stacked FFT blocks and 
a linear network to output a scalar, as shown in Fig. 2(a). 
According to the input character sequence, 
the duration predictor outputs a duration sequence, where each element of the duration sequence represents 
the pronunciation duration of each character. In inference, the output duration sequence is 
used in the length regulator to control the alignment between text and mel-spectrum.

\subsection{Mix Density Network}
\label{ssec:Mix Density Network}


Since the training of Feed-Forward Transformer and the duration predictor requires the correct alignment 
between the text and mel-spectrum of training dataset, 
we design the mix density network \cite{MDNs,MDNforSpeechSynthesis} to learn the alignment. 
As shown in Fig. 2(b), the mix density network  
is composed of multiple stacked linear layers, each followed by 
the layer normalization, ReLU activation and the dropout layer but the last. 
The last linear layer outputs the mean and variance vector of multi-dimensional gaussian distributions, 
which represents the mel-spectrum distribution of each character. 
Note that the mix density network is stacked on top of the FFT blocks on the character side in Feed-Forward Transformer, 
and its input is the same with the length regulator. 
In addition, the mix density network is designed to learn the alignment of training dataset, 
hence it is only used in the model training stage and can be removed in inference. 



\section{Training and Inference}
\label{sec:Training and Inference}


\subsection{Alignment Loss}
\label{ssec:Alignment Loss} 

Inspired by the Baum-Walch algorithm \cite{taylor2009TTSBook,BaumWelch}, 
we propose the alignment loss to train the mix density network 
and learn the correct alignment between text and mel-spectrum.
Let $\boldsymbol{y} = \left\{y_1,y_2, \cdots,y_n \right\}$ denote 
the mel-spectrum sequence, where $n$ is the frame number of the mel-spectrum. And let $\boldsymbol{z}=\left\{z_1, z_2, \cdots, z_m \right\}$ denote the 
multi-dimensional gaussian distribution sequence output from the mix density network, 
where $z_i=(\mu_i,\Sigma_i )$ represents 
mean vector and covariance matrix, and $m$ is equal to the length of input character sequence. We consider the distribution of each dimensional in the 
mel-spectrum is independent, so the covariance matrix $\Sigma_i$  is a diagonal matrix. 
The probability of the $i$-th frame mel-spectrum with respect 
to the $j$-th distribution of the multi-dimensional gaussian distribution sequence 
can be calculated by the multi-dimensional gaussian function:
\begin{equation}
    p\left(y_{i} | z_{j}\right) = \mathcal{N}\left( y_{i} | \mu_j, \Sigma_j \right)
\end{equation}

Let $\ell$ denote the alignment, and $j=\operatorname{Align}(i, \ell)$ denote that the $i$-th 
frame mel-spectrum is aligned to the $j$-th element of the distribution sequence. 
The objective function can be considered as the condition 
probability of the mel-spectrum sequence $\boldsymbol{y}$ with respect to the distribution sequence $\boldsymbol{z}$:
\begin{equation}
    p(\boldsymbol{y}, \ell | \boldsymbol{z} )=\prod_{i=0}^{n} p\left(y_{i} | z_{\operatorname{Align}(i, \ell)}\right)
\end{equation}

However, due to the unknown of the alignment, we cannot directly calculate the above equation. 
Fortunately, the Baum-Welch algorithm provides a great idea to solve this problem, which considers 
all possible alignment and sums their probability. 
Therefore, the objective function is designed as
\begin{equation}
    p(\boldsymbol{y} | \boldsymbol{z})=\sum_{\ell} \prod_{i=0}^{n} p\left(y_{i} | z_{\text {Align}(i, \ell)}\right)
\end{equation}

Finally, the alignment loss is designed as
\begin{equation}
    \mathcal{L}_{\text{align}}(\boldsymbol{z}, \boldsymbol{y})=-\log p(\boldsymbol{y} | \boldsymbol{z})
\end{equation} 

When the alignment loss converges, 
the correct alignment can be extracted using the Viterbi algorithm \cite{taylor2009TTSBook}.

\subsection{Training} 
\label{ssec:Training} 



\subsubsection{Forward Algorithm for Alignment Loss} 
\label{sssec:Forward algorithm}

Similar to the forward-backward algorithm for HMMs, 
we apply a dynamic programming algorithm to efficiently calculate the loss. 
Define the forward variables as 
\begin{equation}
    \alpha_{t, s}=p\left(\boldsymbol{y}_{1 : t} | \boldsymbol{z}_{1 : s}\right)
\end{equation}
\noindent where $\boldsymbol{y}_{1:t}=\left\{y_1,y_2, \cdots,y_t \right\}, 0 < t \leq  n$ and 
$\boldsymbol{z}_{1:s}=\left\{z_1, z_2, \cdots, z_s \right\}, 0 < s  \leq m$. 
The following recursion is established 
\begin{equation}
    \alpha_{1,1} = p\left(y_{1} | z_{1}\right),\quad  \alpha_{1, s} = 0 \quad \forall 1 < s \leq m
\end{equation}

\noindent and $\forall 1<t \leq n, 1 \leq s \leq m$, 
\begin{equation}
    \alpha_{t, s}=\left(\alpha_{t-1, s}+\alpha_{t-1, s-1}\right) \cdot p\left(y_{t} | z_{s}\right)
\end{equation}

Then the alignment loss can be calculated by the above recursive formula:
\begin{equation}
    \mathcal{L}_{\text{align}}(\boldsymbol{z}, \boldsymbol{y}) 
    = -\log p\left(\boldsymbol{y}_{1 : n} ; \boldsymbol{z}_{1 : m}\right) 
    = -\log \alpha_{n, m}
\end{equation}

Note that the alignment loss is differentiable with respect to the distribution sequence $\boldsymbol{z}$ since its just sums, 
products and exponents of them. Therefore, the mix density network can be trained using the gradient descent 
algorithm applied by Tensorflow or Pytorch. 





\subsubsection{Multi-phases Training} 
\label{sssec:Multi-phases Training}

Since the training of Feed-Forward Transformer and the duration predictor 
requires the correct alignments which are obtained from the trained mix density network, 
it is difficult to jointly train the entire network at once in practice. 
Therefore, we adopt multi-phases training for AlignTTS. 
\begin{itemize}
  \item Firstly, we train the mix density network with the first FFT blocks of 
  the Feed-Forward Transformer using the alignment loss.
  \item Secondly, we extract the alignment and convert it into the duration sequence. 
  Fixing the parameters of the first FFT blocks, 
  the rest network of Feed-Forward Transformer is trained 
  using the mean square error (MSE) loss between the predicted and target mel-spectrum.
  \item Then, the Feed-Forward Transformer and the mix density network are trained together to fine-tune parameters, 
  where the duration sequence is calculated in each training step. 
  \item Finally, we use the final mix density network to calculate the character durations 
  and train the duration predictor using the MSE loss in logarithmic domain \cite{FastSpeech}. 
\end{itemize}

\subsection{Inference}
\label{ssec:Inference}

In inference, input characters are convert to $N$-dimensional character embedding sequence which are passed 
through the first FFT blocks of Feed-Forward Transformer to output the hidden feature sequence. 
At the same time, the duration predictor generates the duration sequence, which is used by the length regulator 
to expand the hidden feature sequence. The rest network of Feed-Forward Transformer predicts the mel-spectrum 
based on the expanded hidden feature sequence. 
Finally, a parallel neural vocoder, 
such as WaveGlow, is used to generate waveform according to
the predicted mel-spectrum.  


\section{Experiments}
\label{sec:experiment}

\subsection{Datasets}
\label{ssec:datasets} 

We train AlignTTS on LJSpeech dataset \cite{ljspeech17}, 
which is randomly divided into 2 sets: 12600 samples for training, and 500 samples for testing. 
The computation of the mel-spectrum is the same with the description in WaveGlow \cite{WaveGlow}. 
The texts are normalized and inserted with 
the beginning character (a space) and the end character (a period), e.g. ``There are 16 apples'' is 
converted to `` there are sixteen apples.''. 

\subsection{Model Configuration}
\label{ssec:model configuration}

The Feed-Forward Transformer contains 6 FFT blocks on both the character side and the mel-spectrum side, 
and the duration predictor includes 2 FFT blocks. 
The dimension of each network in the Feed-Forward Transformer is all set to 768 
and the dimension in the duration predictor is all set to 128.
The number of attention head is set to 2 and the kernel size of 1D convolution is set to 3 in all FFT block.
The hidden size of the linear layer in the mix network is set to 256 and the dimension of the output is 
160 (80 dimensions for the mean and 80 dimensions for variance of the gaussian distribution). 

We train our model on 2 NVIDIA V100 GPUs, with batch size of 16 samples on each GPU, and use the Adam optimizer 
with $\beta_1=0.9$, $\beta_2=0.98$, $\varepsilon=10^{-9}$. 
we adopt the same learning rate schedule in \cite{AttentionAllYouNeed} 
with 40K training steps in the first two training stages, 
and a fixed learning rate of $10^{-4}$ with 80K training steps
in fine-tuning the parameters of the whole model. 
In addition, the duration predictor is trained with a fixed learning rate of $10^{-4}$ and 10K training steps.

\subsection{Evaluation}
\label{ssec:Evaluation}

In order to evaluate the performance of our model, we use the text transcripts 
in test datasets as the input of the model, 
and obtain the synthetic audios, which are rated together 
with the ground truth audio (GT) by 50 native English speakers. 
And then the mean opinion score (MOS) is calculated as 
the evaluation indicator of the text-to-speech systems. 
As comparative experiments, the audios generated from mel-spectrum of the ground truth audio (GT Mel), 
Tacotron2 \cite{Tacotron2}, Transformer TTS \cite{TransformerTTS} and 
FastSpeech \cite{FastSpeech} are also rated together. 
The vocoder is implemented using the WaveGlow \cite{WaveGlow} in these experiments. 
The results are shown in Table 1, 
where the time cost for generating one speech is also illustrated. 
It can be seen that our proposed model gets 
slightly improved performance than Tacotron2 and Transformer TTS, 
and a significant speed boost is acquired in inference. 
Specifically, it takes only 0.18 seconds to synthesize approximately 10 seconds of speech in our model, 
of which about 0.06 seconds for AlignTTS and about 0.12 seconds for WaveGlow.

\begin{table}[t]
  \caption{The comparison of MOS and time cost.}
  \begin{center}
  \begin{tabular}{p{3cm}p{2cm}<{\centering}p{2cm}<{\centering}}
  \toprule
  \textbf{Method}&\textbf{MOS}&\textbf{Time Cost} (s) \\
  \midrule
  GT & $4.53 \pm 0.05$ & - \\
  GT Mel & $4.28 \pm 0.08$ & $0.12 \pm 0.02$ \\
  Tacotron2 & $3.96 \pm 0.13$ & $4.58 \pm 2.87$ \\
  Transformer TTS  & $4.02 \pm 0.15$ & $3.26 \pm 1.93$ \\
  FastSpeech  & $3.88 \pm 0.11$ & $ 0.16 \pm 0.04 $ \\
  \midrule
  AlignTTS & $ \mathbf{4.05 \pm 0.12}$ & $0.18 \pm 0.04$ \\
  \bottomrule
  \end{tabular}
  \label{Table1}
  \end{center}
\end{table}

\subsection{Comparison of Alignment}
\label{ssec:Comparison of Alignment}

We design the alignment loss to learn the alignment, while FastSpeech extract it from Transformer TTS 
model. Comparison of the alignment between our model and FastSpeech is shown in Figure 3. We can find that 
the alignment from our model is more precise in aligning characters and mel-spectrum, 
which effectively guides the training and 
makes model more excellent and robust than other TTS systems.

\begin{figure}[t]
  \centering
  \includegraphics[width=0.9\linewidth]{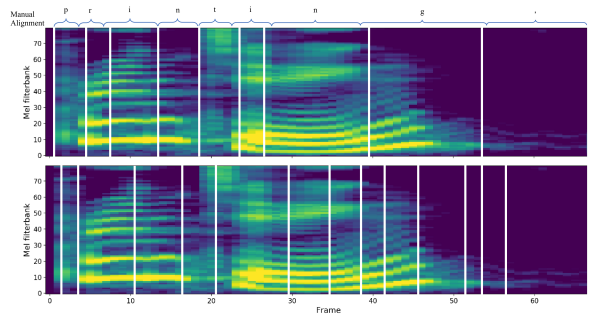}
  \caption{ Comparison of Alignments. 
  The mel-spectrum is split by the alignment of 
  AlignTTS (above) and FastSpeech (below), respectively. 
  The corresponding text is `` printing, ...''. 
    }
  \label{Figure3}
\end{figure}

\section{Conclusion} 

In this work, we propose AlignTTS to predict the mel-spectrum in parallel, 
and the alignment loss to make model capable
of learning the alignment between text and mel-spectrum. 
Experiments on the LJSpeech dataset show 
that the more precise alignment is obtained, 
which allows AlignTTS to better learn the transformation from characters to mel-spectrum, 
and the state-of-the-art performance is achieved. 

\section{Acknowledgements}

This paper is supported by National Key Research and Development Program of China under grant No. 2018YFB1003500, No. 2018YFB0204400 and No. 2017YFB1401202. 
Corresponding author is Jianzong Wang from Ping An Technology (Shenzhen) Co., Ltd.


\bibliographystyle{IEEEbib}

\end{document}